\documentclass[aps,prb,reprint,floatfix]{revtex4-2}
\usepackage[T1]{fontenc}
\usepackage[utf8]{inputenc}

\usepackage{graphicx}

\usepackage[dvipsnames]{xcolor}
\usepackage[
colorlinks=true,allcolors=Blue,%
pdfusetitle,%
]{hyperref}

\usepackage{amsmath,amssymb,mathtools}
\usepackage{bm}
\usepackage{siunitx}

\AtBeginDocument{
\newcommand{\dif}{\mathrm{d}}
\newcommand{\E}{\mathrm{e}}
\newcommand{\I}{\mathrm{i}}
\let\Re\relax \let\Im\relax
\DeclareMathOperator{\Re}{Re}
\DeclareMathOperator{\Im}{Im}
}

\begin{document}
\title{Geometric effects in the Dyakonov--Shur theory of Teraherz photodetection}
\author{Riccardo Riolo}
\affiliation{Dipartimento di Fisica dell'Universit\`a di Pisa, Largo Bruno Pontecorvo 3, I-56127 Pisa, Italy}
\author{Marco Polini}
\affiliation{Dipartimento di Fisica dell'Universit\`a di Pisa, Largo Bruno Pontecorvo 3, I-56127 Pisa, Italy}
\author{Riccardo Mannella}
\affiliation{Dipartimento di Fisica dell'Universit\`a di Pisa, Largo Bruno Pontecorvo 3, I-56127 Pisa, Italy}
\author{Andrea Tomadin}
\affiliation{Dipartimento di Fisica dell'Universit\`a di Pisa, Largo Bruno Pontecorvo 3, I-56127 Pisa, Italy}
%


\begin{abstract}
Nonlinear resonances of plasma waves in field-effect transitors
enable a well-known photodetection mechanism,
first introduced by Dyakonov and Shur in the Nineties,
especially suited to the Terahertz (THz) frequency range.
Theoretical analyses of the mechanism always assume that the
gate of the transistor, which is coupled to the antenna receiving the THz signal,
is as long as the channel itself, at odds with typical experimental devices,
where short gates are usually employed, e.g.\ due to fabrication constraints.
In this work we overcome this limitation and provide a complete theory
of Dyakonov--Shur photodetection in the presence of short gates.
We develop our theory in such a general fashion that allows us to also
treat the case in which multiple gates are present.
We find that a sub-optimal positioning of the gates can substantially
decrease the detection efficiency of the device and provide a compact
analytical formula to quickly estimate the optimal gate position.
\end{abstract}

\maketitle

\section{Introduction}
Nonlinear effects in photonics and opto-electronics have been
captivating a great deal of attention for decades~\cite{Boyd,Shen}.
Typically, these nonlinearities stem from the intrinsic crystalline
structure of the material under investigation, which determine its
nonlinear polarization functions such as the second- and third-order
ones. Another more subtle source of nonlinearity, however, stems from
the motion of electrons in a crystal, in the so-called hydrodynamic
transport regime~\cite{Polini_2020}. As in classical fluid dynamics,
these hydrodynamic nonlinearities are convection and the coupling of two
collective variables, i.e.\ density and fluid-element velocity.
Hydrodynamic effects in a system of electrons roaming in a crystal arise
when electron-electron collisions take place on a time scale
$\tau_{\text{ee}}$ that is much shorter than the time scale of the
evolution of the above mentioned collective quantities.
As the electrons move at the Fermi velocity $v_{\text{F}}$,
the mean free path for electron-electron collisions is
$\ell_{\text{ee}} = v_{\text{F}}\tau_{\text{ee}}$.
In order to apply the hydrodynamic theory,
$\ell_{\text{ee}}$ must be the shortest length scale in the system.
In particular $\ell_{\text{ee}}$ must be shorter than the length $L$ of
the sample and shorter than the mean free path of electrons,
determined by the scattering rate $1/\tau$ for
collisions with phonons, impurities and defects,
which do not conserve momentum.

In recent years, the realization of the hydrodynamic regime has been
demonstrated in pure two-dimensional (2D) electron systems, where the
parameters which control the time- and length-scales mentioned above can
be easily tuned with the aid of metallic gates. Examples include
graphene~\cite{Bandurin_2016,Crossno_2016,Kumar_2017,Gallagher_2019,%
Berdyugin_2019,Sulpizio_2019,Ku_2020,Kumar_2022},
GaAs/AlGaAs quantum wells~\cite{Keser_2021,Ginzburg_2021,Wang_2022},
and the highly-pure Weyl semimetal $\mathrm{WTe}_2$~\cite{Steinberg_2022}.

The theory of hydrodynamic transport in graphene~\cite{%
Torre_2015,Narozhny_2015,Levitov_2016,Narozhny_2017,Guo_2017,Shytov_2018,Ledwith_2019}
shows some peculiarities due to the linear electron dispersion
in the vicinity of the charge neutrality point,
and changes qualitatively when the Fermi energy is smaller
than the energy scale associated to the
electron temperature~\cite{Fritz_2008, Muller_2009}.
The scales of interest for the applicability of the hydrodynamic theory
strongly depend on the density $n$ and temperature $T$ of the electron liquid.
For both GaAs~\cite{Giuliani_2005}
and single-layer graphene~\cite{Li_2013, Polini_2016},
the inverse lifetime of a quasiparticle of energy $\varepsilon_k$, in the limit
$\lvert\varepsilon_k-\varepsilon_{\text{F}}\rvert\ll k_{\text{B}}T\ll \varepsilon_{\text{F}}$,
$\varepsilon_{\text{F}}$ being the Fermi energy, 
scales with temperature as $1/\tau_{\text{ee}}\propto-T^2\ln T$.

What are the implications of hydrodynamic electron transport on the
fabrication of a device of some use in applications? 
The first device of practical interest was proposed theoretically by
Dyakonov and Shur (DS) in the Nineties~\cite{%
Dyakonov_1993,Dyakonov_1995,Dyakonov_1996a,Dyakonov_1996b}.
DS considered a field-effect transistor (FET) where the 2D
electron fluid in the channel is in the hydrodynamic transport regime.
This device can be used as a THz photodetector~\cite{Dyakonov_1996a}
by connecting an antenna to the source and gate contacts.
Indeed, a DC potential (the \emph{photovoltage}) is established between
source and drain when the FET is driven by an AC signal, which is thus
\emph{rectified}.
The underlying mechanism involves the generation and propagation of
electron density waves ({\it plasma waves})~\cite{%
Dyakonov_1993,Dyakonov_1995,Dyakonov_1996a,%
Dyakonov_1996b,Tomadin_2013,Tomadin_2021}, which mix because
of the hydrodynamic nonlinearities.
For the plasma waves to be described by the hydrodynamic model, however,
their frequency $\omega$ must be much shorter than the electron-electron
collision rate $1/\tau_{\text{ee}}$.
This ``collisional'' regime is the opposite of the usual
``collisionless'' regime of plasmons~\cite{Giuliani_2005},
where electrons can be understood to move freely under the action of a
collective field.
For GaAs at $n=\SI{e11}{cm^{-2}}$, $T=\SI{10}{K}$,
the scattering length is $\ell_{\text{ee}}\sim\SI{200}{nm}$
and the maximum frequency of plasma waves
$f_{\text{max}}\sim\SI{100}{GHz}$.
On the other hand, for single-layer graphene (SLG) at $n=\SI{e12}{cm^{-2}}$,
$T=\SI{200}{K}$, they are $\ell_{\text{ee}}\sim\SI{200}{nm}$ and
$f_{\text{max}}\sim\SI{1}{THz}$, respectively.
DS theory predicts a frequency-resolved detection for a sufficiently
clean channel, i.e.\ when the frequency $\omega$ is larger
than the scattering rate $1/\tau$.

The DS rectification mechanism is of practical importance in the detection of
electromagnetic radiation in the THz range because frequencies in this
range exceed the cutoff frequency of electronics circuitry,
and other devices based on e.g.\ Schottky barriers, photothermoelectric or photogalvanic
effects~\cite{Koppens_2014} offer limited bandwidth or need cryogenic temperatures.
The original theory of DS detection dealt with a parabolic-band
2D electron system as the one that can be found in a GaAs electron-doped quantum well but has been extended to
devices based on SLG~\cite{Tomadin_2013}
and bilayer graphene~\cite{Tomadin_2021}, where plasma waves at typical
electron densities fall naturally in the THz range and ultra-high
electron mobilities at room temperature
reduce the minimum frequency necessary for frequency-resolved detection.
Similarly, experiments in DS photodetection first used FETs
based on GaAs heterostructures~\cite{Knap_2009}
and then graphene-based devices ~\cite{Vicarelli_2012,
Spirito_2014, Tredicucci_2014}. Resonant, frequency-resolved
detection was finally measured in bilayer graphene in 2018 in the
pioneering experiments by Bandurin et al.~\cite{Bandurin_2018}.

Notwithstanding this multi-decade-long research activity in DS photodetection,
most theoretical works deal with very simple geometries,
where a single gate occupies the entire length of the FET channel.
However, in a typical experimental setup, the width of the employed
gates often tends to be much shorter than the channel
length~\cite{Vicarelli_2012,Spirito_2014} due to fabrication constraints.
As a consequence, the position of the gate between source and drain becomes
a new geometrical parameter that has to be determined before fabrication
on the basis of some criterion
(for example, in order to maximize the photodetector efficiency).
In this work, we systematically address for the first time
how the gate length and its precise positioning affect the photovoltage.
Moreover, motivated by current fabrication possibilities, we additionally study
the effects of \emph{multiple} gates in a single FET.

Our Article is organized as following.
In Sec.~\ref{sec:hydro}, we review the theory of electron
hydrodynamics for both GaAs and SLG.
In Sec.~\ref{sec:geom}, we discuss the electrostatics of a device with
an arbitrary number of top gates and the boundary conditions needed in
order to operate as a detector.
In Sec.~\ref{sec:resp}, we calculate the photovoltage.
We discuss the plasma-waves solution of the hydrodynamic equations
and then the second-order solution leading to the DC signal.
Moreover, we provide an approximate formula in the limit of a single
gate of negligible width.
Finally, a summary of our main results and a brief set of conclusions
are reported in Sec.~\ref{sec:results}.

\section{Brief recapitulation of hydrodynamic theory}
\label{sec:hydro}

In the hydrodynamic regime,
electron-electron scattering is the process
which is responsible for the local equilibration of the electron system.
Within the kinetic theory approach~\cite{Landau:10},
the electron system can be described by the 
distribution function $f_{\lambda \bm k}(\bm{r},t)$,
which measures the probability that an electron with wave vector
$\bm{k}$ and belonging to  band $\lambda$ is found 
in the neighborhood of position $\bm{r}$ at time $t$.
The explicit form of the probability distribution function
in a state of quasi-equilibrium (i.e.~when thermodynamic
quantities are locally well-defined, but might vary in
space and time)
is obtained by requiring that that the collisional
integral for electron-electron scattering in the Boltzmann equation
vanishes~\cite{Rudin_2011,
Svintsov_2012, Narozhny_2022}.
Assuming that the electron dispersion is isotropic,
the probability distribution depends only on the magnitude $k$
of the wave vector or, equivalently, on 
the single-particle state energy $\varepsilon_{\lambda k}$.
In the following, we will also assume that the 2D system is
inhomogeneous in the $x$ direction only (this will be the
source-drain direction when applying the theory to a FET in the
DS setup).
The result is that $f_{\lambda k}$ is a drifted Fermi--Dirac distribution,
\begin{equation}\label{eq:fermidirac}
f_{\lambda k}(x,t) = \left ( \exp\frac{\varepsilon_{\lambda k}-\hbar v(x,t) k
- \mu(x,t)}{k_{\text{B}} T} + 1 \right )^{-1},
\end{equation}
where $v(x,t)$ is the local drift velocity,
$\mu(x,t)$ the local chemical potential,
and $T$ the temperature.
We shall assume throughout this work that the temperature is much smaller than the
Fermi temperature, i.e.~$T\ll T_{\text{F}}$.
In such limit, the distribution~\eqref{eq:fermidirac}
becomes the zero-temperature Fermi--Dirac distribution.
Thus there is a local Fermi energy $\varepsilon_{\text{F}}(x,t)$,
with a Fermi wavevector $k_{\text{F}}(x,t)$ that smoothly depends on $x$
and $t$ through the local density $n(x,t)$,
\begin{equation}
k_{\text{F}}(x,t) = \sqrt{\frac{4\pi n(x,t)}{g}},
\end{equation}
where $g$ is the degeneracy of single-particle states.

The hydrodynamic equations are obtained from the Boltzmann kinetic
equation by considering the moments of the distribution function,
i.e.\ the density and drift velocity
\begin{gather}
n_{\lambda}(x,t)\equiv \frac{1}{L^2}\sum_{\bm k}f_{\lambda\bm k}(x,t) \\
{\bm v}_{\lambda}(x,t)\equiv \frac{1}{L^2}\sum_{\bm k}
f_{\lambda\bm k}(x,t) \nabla_{\bm k}\varepsilon_{\lambda\bm k}.
\end{gather}

The first hydrodynamic equation is the continuity equation,
which expresses the conservation of the particle number.
The second hydrodynamic equation is the Navier--Stokes equation,
which expresses the conservation of momentum.
By assuming a uniform temperature $T$, there is no heat transport,
therefore we can neglect the thermal transport equation,
which expresses the conservation of energy.
The actual form of the hydrodynamic equations depends on the electronic
dispersion in the material that hosts the 2D electron gas, and has
been discussed e.g.\ in Ref.~\cite{Narozhny_2022}.

Dissipative effects in hydrodynamic theory are characterized by
three transport coefficients~\cite{Landau:6, Landau:10}:
the shear viscosity, which describes the friction between layers of
fluid moving with different velocities;
the bulk viscosity, which describes the dissipation due to a
compression or expansion of the liquid;
and the thermal conductivity, which describes the dissipation
of energy due to temperature gradients.
In the low-frequency hydrodynamic limit, the bulk viscosity
vanishes~\cite{Giuliani_2005, Principi_2016}, while
the kinematic shear viscosity for both GaAs~\cite{Giuliani_2005} and
grafene~\cite{Principi_2016} is $\nu=v_{\text{F}}\ell_{\text{ee}}/4$.
The interactions of the electrons with phonons or impurities,
which tend to restore the equilibrium distribution $f_k^{({\rm eq})}$,
are included in the theory with the so-called relaxation-time approximation,
which adds the term $-(f_k(x,t) - f_k^{({\rm eq})})/\tau$
to the Navier-Stokes equation.
Here, the phenomenological parameter $\tau$  represents the
characteristic time scale over which the momentum of a fluid element is
randomized.

The hydrodynamic equations relate the density $n$,
the drift velocity $v$, the electrostatic potential $\varphi$
and the pressure of the electron gas $P$.
Therefore, alongside these two equations,
we need two constitutive relations:
the solution to the electrostatic problem,
which provides the relation $\varphi(n)$,
and the equation of state $P(n)$,
which is dependent on the electron dispersion.
We now report the hydrodynamic equations for both 2D
parabolic-band electron systems (GaAs being the cleanest example in
terms of electronic mobility) and SLG,
following Refs.~\cite{Rudin_2011, Svintsov_2012, Narozhny_2022}.

\subsection{Gallium arsenide}
Conduction-band electrons in $n$-doped GaAs have a parabolic dispersion
\begin{equation}
\varepsilon_k = \frac{\hbar^2 k^2}{2m},
\end{equation}
where $m$ is the band mass, whose value for GaAs is
$m\approx 0.063m_{\text{e}}$,
where $m_{\text{e}}$ is the bare electron mass in vacuum.
The degeneracy factor is $g=2$ and is solely due to the electron's spin.

The continuity equation reads as following,
\begin{equation}\label{eq:contgaas}
\partial_t n + \partial_x(nv) = 0,
\end{equation}
while the Navier--Stokes equation is 
\begin{equation}
\partial_t v + v\partial_x v
= \frac{e\partial_x\varphi(n)}{m} - \frac{\partial_x P(n)}{nm}
- \frac{v}{\tau} + \nu\partial_x^2 v,
\end{equation}
where $\tau$ is the momentum relaxation time
and $\nu$ the kinematic viscosity.
In the limit $T\ll T_{\text{F}}$, the equation of state reads
\begin{equation}\label{eq:pgaas}
P(n) = \frac{n \varepsilon_{\text{F}}(n)}{2}
= \frac{\hbar^2\pi n^2}{2m}.
\end{equation}
Inserting this result into the Navier--Stokes equation we find
\begin{equation}\label{eq:eulergaas}
\partial_t v + v\partial_x v
= \frac{e\partial_x\varphi(n)}{m} - \frac{\pi\hbar^2\partial_x n}{m^2}
- \frac{v}{\tau} + \nu\partial_x^2 v.
\end{equation}

\subsection{Single-layer graphene}
Electrons in SLG, in the vicinity of the charge neutrality point, have a linear dispersion
\begin{equation}
\varepsilon_{\lambda k} = \lambda \hbar v_{\text{F}} k,
\end{equation}
where $v_{\text{F}}\approx\SI{e8}{cm/s}$ is the Fermi velocity and
$\lambda=\pm$ for the conduction and valence band, respectively~\cite{Katsnelson_2020}.
Twofold spin and valley degeneracies yield $g=4$.

The continuity equation takes the same form as in GaAs
\begin{equation}\label{eq:contslg}
\partial_t n + \partial_x(nv) = 0,
\end{equation}
where $n(x,t)=n_+(x,t) + n_-(x,t)$ is total number density.
In the limit in which the drift velocity is much smaller than the Fermi
velocity, $v(x,t)\ll v_{\text{F}}$, the Navier--Stokes equation
is~\cite{Tomadin_2013}
\begin{multline}
\frac{3P}{v^2_{\text{F}}}
\bigg(\partial_t v + v\partial_x v
- \frac{v\partial_t n}{n}
+ \frac{v}{\tau} - \nu\partial_x^2 v\bigg)
+ \frac{3v}{v^2_{\text{F}}}\partial_t P \\
= -\partial_x P + (n_+ - n_-)e\partial_x \varphi(n).
\end{multline}
In the limit $T\ll T_{\text{F}}$, the equation of state reads
\begin{equation}\label{eq:pslg}
	P(n) = \frac{n \varepsilon_{\text{F}}(n)}{3}
	= \frac{\pi^{1/2}\hbar v_{\text{F}}n^{3/2}}{3}.
\end{equation}
Away from the charge neutrality point,
i.e.\ for a small hole population $n_-\ll n_+\approx n$,
the Navier--Stokes equation reduces to the simpler form
\begin{equation}\label{eq:eulerslg}
\partial_t v + v\partial_x v
= \frac{v_{\text{F}}^2e\partial_x\varphi(n)}{\varepsilon_{\text{F}}(n)}
- \frac{v_{\text{F}}^2\partial_x n}{2n}
- \frac{v\partial_t n}{2n}
- \frac{v}{\tau} + \nu\partial_x^2 v.
\end{equation}

\section{Geometry of the device}\label{sec:geom}
\subsection{Electrostatics of a single-gated and dual-gated transistor}
\begin{figure}
\centering
\includegraphics{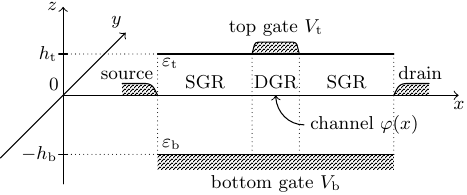}
\caption{%
Schematics of a FET with a single top gate.
The channel lies on the plane $z=0$ and the system is translationally invariant along the $y$ direction.
The dielectric constant is $\varepsilon_{\text{t}}$ ($\varepsilon_{\text{b}}$)
and the gate layer thickness is $h_{\text{t}}$ ($h_{\text{b}}$) above (below) the channel.
The channel region where both the top and the bottom gates are present
is denoted dual-gated region (DGR) while
the rest are denoted single-gated regions (SGRs).
The electric potential in the channel, top, and bottom gate is
$\varphi(x)$, $V_{\text{t}}$, and $V_{\text{b}}$, respectively.
}\label{fig:fet1}
\end{figure}

The FET channel that hosts the 2D electron liquid
lies in the plane $z=0$
and extends along the $x$ direction for a total length $L$.
Throughout this work we will assume that the device is translationally invariant
in the $y$ direction.
The source and drain contacts are connected
to the two extrema of the channel,
as shown schematically in Fig.~\ref{fig:fet1}.
We model the top and bottom gate as two perfect conductors lying in the planes $z=h_{\text{t}}$
and $z=-h_{\text{b}}$, respectively.
The dielectric constant of the medium surrounding the channel
is $\varepsilon_{\text{t}}$ for $0<z<h_{\text{t}}$
and $\varepsilon_{\text{b}}$ for $-h_{\text{b}}<z<0$.
External voltage generators fix the electric potential
of the top and bottom gates to the values $V_{\text{t}}$ and $V_{\text{b}}$,
respectively, while the eletric potential $\varphi(x)$ in the channel
is not homogeneous and is determined by the hydrodynamics and constitutive equations,
as discussed above.

As discussed in the Introduction, in this work
we consider a device in which the top gate only
partially extends over the channel.
The bottom gate, on the contrary, extends over the whole
channel from source to drain.
With this configuration in mind,
we distinguish between the regions of the $x$ axis where 
only the bottom gate is present,
called {\it single-gated regions} (SGRs),
and the regions where both gates are present,
called {\it dual-gated regions} (DGRs).

If the distances $h_{\text{t}}$, $h_{\text{b}}$
between the gates and the channel are
much smaller than the length scale on which the electron density varies,
then the relation between the electric potential and the density assumes
a local form.
In this {\it local capacitance approximation}~\cite{Dyakonov_1993,
Dyakonov_1995, Dyakonov_1996a, Dyakonov_1996b, Tomadin_2013,
Tomadin_2021}, the electric field is directed along the $z$ direction, $E_z(x,z)$.
(Here we neglect the fringe field due to the finite dimension of
the gate conductors.)
The electric displacement field is $D_z(x,z) = \varepsilon(z) E_z(x,z)$.
From Gauss' law $\partial_z D_z(x,z) = -4\pi e \delta(z) n(x)$ we find
the relation between the displacement field on opposite sides of the
channel
\begin{equation}
D_z(x,z=0^+) - D_z(x,z=0^-) = -4\pi en(x).
\end{equation}
The electric displacement field above the channel depends on whether $x$ is
in the SGRs or in the DGR,
\begin{subequations}
\begin{equation}
D_z(x,z=0^+) = \begin{dcases}
0, & \text{if $x \in \text{SGRs}$}, \\
-\frac{\varepsilon_{\text{t}}(V_{\text{t}}-\varphi(x))}{h_{\text{t}}},
& \text{if $x \in \text{DGRs}$},
\end{dcases}
\end{equation}
while the field below the channel is the same in every region
\begin{equation}
D_z(x,z=0^-)
= \frac{\varepsilon_{\text{b}}(V_{\text{b}}-\varphi(x))}{h_{\text{b}}}.
\end{equation}
\end{subequations}
The relation between the electron density in the channel and the
electric potentials depends on the region,
\begin{multline}\label{eq:localcapac}
en(x) \\
= \begin{cases}
C_{\text{b}}(V_{\text{b}}-\varphi(x)),
& \text{if $x \in \text{SGRs}$}, \\
C_{\text{b}}(V_{\text{b}}-\varphi(x))
+ C_{\text{t}}(V_{\text{t}}-\varphi(x)),
& \text{if $x \in \text{DGRs}$},
\end{cases}
\end{multline}
where $C_{\text{t}}$ and $C_{\text{b}}$ are the geometrical capacitances
per unit area associated to top and bottom gate, respectively, given by
\begin{equation}
C_{\text{t,b}} = \frac{\varepsilon_{\text{t,b}}}{4\pi
h_{\text{t,b}}}.
\end{equation}

\begin{figure}
\centering
\includegraphics{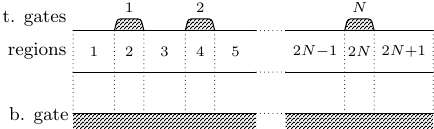}
\caption{%
Schematics of a FET configuration with $N$ top gates.
We number the regions below and between the gates
with an index $j = 1, \dots, 2N+1$,
so that the regions with $j$ odd (even) are SGRs (DGRs).
}\label{fig:fet2}
\end{figure}

Finally, we generalize the previous discussion to a geometry
with several top gates, as shown in Fig.~\ref{fig:fet2}.
Let us consider the case of $N$ top gates,
which divide the channel into $2N+1$ regions,
each of length $\ell_j$, with $j=1,2,\dots,2N+1$,
so that $\sum_j \ell_j=L$.
We denote the density and electric potential in the $j$-th region by
$n^{(j)}(x,t)$ and $\varphi^{(j)}(x,t)$, respectively, where the spatial coordinate $x$ varies in the interval $(0,\ell_j)$.
With this numbering convention,
the {\it odd} regions are SGRs,
while the {\it even} regions are DGRs.
The average electron density $n_0$ in the channel
is determined by the voltage applied to the bottom gate,
which occupies the whole length of the device,
according to the relation $n_0=C_{\text{b}}V_{\text{b}}/e$.

\subsection{Dyakonov--Shur boundary conditions}
The \unit{THz} radiation, collected by the lobes of the antenna,
is funneled into the FET contacts
and determines the electric potential difference
between the source contact and the top gates.
For definiteness, we assume that the source contact
is grounded and that the voltage $V_{t}(t)$
common to all the top gates
oscillates with the frequency $\omega$
of the incoming radiation and with
an amplitude $U_{\text{a}}$, which depends on
implementation details such as the efficiency of the antenna
and its impedence matching with the FET.
Moreover, the DS scheme requires that the drain contact
is floating, so that the current density $j(x,t) = n(x,t) v(x,t)$
vanishes at $x = L$.
These requirements provide the following Dyakonov--Shur boundary conditions
that complement the hydrodynamic and constitutive equations
discussed above:
\begin{subequations}
\begin{gather}\label{eq:boundary}
\varphi^{(1)}(0,t) = 0, \\
V_{\text{t}}(t) = U_{\text{a}}\cos(\omega t), \\
j^{(2N+1)}(\ell_{2N+1},t) = 0.
\end{gather}
\end{subequations}
\section{Calculation of the photovoltage}\label{sec:resp}

\subsection{Perturbative expansion}
The photovoltage is the DC component of the
electric potential difference between source and drain
\begin{equation}\label{eq:dudef}
\Delta U = \big\langle\varphi^{(1)}(0,t)
- \varphi^{(2N+1)}(\ell_{2N+1},t)\big\rangle_t,
\end{equation}
where $\langle a(t)\rangle_t$ denotes the time average of the oscillating
signal $a(t)$,
\begin{equation}
\langle a(t)\rangle_t \equiv \frac{\omega}{2\pi}\int_0^{2\pi/\omega}
a(t)\dif t,
\end{equation}
where $\omega$ is its angular frequency.
Since the full hydrodynamic equations are not amenable to an analytical solution,
to calculate the photovoltage we follow the DS procedure~\cite{Dyakonov_1996a}, which 
consists in a perturbative expansion up to the second order
in the amplitude $U_{\rm a}$ of the voltage oscillation
produced by the incoming radiation.

For both GaAs and SLG,
taking into account the boundary conditions in Eq.~\eqref{eq:boundary},
the steady solution of the hydrodynamic equations is
\begin{equation}\label{eq:steady}
v(x,t) = 0, \quad
\varphi(x,t) = 0, \quad
n(x,t) = n_0.
\end{equation}

Due to the nonlinearities in Eqs.~(\ref{eq:contgaas}), (\ref{eq:eulergaas}),
(\ref{eq:contslg}), and~(\ref{eq:eulerslg}),
a perturbative, monochromatic drive of frequency $\omega$ generates
oscillations at all the harmonics of $\omega$, including
a DC component at zero frequency.
Expanding around the steady state defined by Eq.~\eqref{eq:steady},
up to the second order in the amplitude $U_{\text{a}}$,
we seek solutions of the form
\begin{subequations}
\begin{align}
v(x,t) &= v_1(x,t) + v_2(x,t) + \delta v(x), \\
\varphi(x,t) &= \varphi_1(x,t) + \varphi_2(x,t) + \delta \varphi(x), \\
n(x,t) &= n_0 + n_1(x,t) + n_2(x,t) + \delta n(x),
\end{align}
\end{subequations}
where all the quantities with subscript $l$ oscillate at frequency
$l\omega$ and are proportional to $(U_{\text{a}})^l$,
while $\delta v$, $\delta \varphi$, $\delta n$ are the DC
components of the second order.
This expansion leads to a set of linear equations of the first order and
a set of the second order, which depends on the solution of the former.
Therefore, in order to calculate the photovoltage $\Delta U$,
i.e.\ the DC component of the second order,
we now proceed to solve the first-order equations.

\subsection{First-order solution: plasma waves}
The local capacitance approximation in Eq.~\eqref{eq:localcapac}, at the first
order in $U_{\text{a}}$, becomes
\begin{equation}\label{eq:localcapac1}
en_1^{(j)} = \begin{cases}
-C_{\text{b}}\varphi_1^{(j)}, & \text{for~odd $j$}, \\
-(C_{\text{b}}+C_{\text{t}})\varphi_1^{(j)}
+U_{\text{a}}\cos\omega t, & \text{for~even $j$}.
\end{cases}
\end{equation}
The first-order hydrodynamic equations are equal in form
for both GaAs and SLG
\begin{gather}
\partial_t n_1 + n_0\partial_x v_1 = 0,
\label{eq:cont1} \\
\partial_t v_1
= \frac{e}{m}\partial_x\bigg(
\varphi_1 - \frac{en_1}{C_{\text{q}}}\bigg)
- \frac{v_1}{\tau} + \nu\partial_x^2 v_1.
\label{eq:euler1}
\end{gather}
In the last equation, the effective mass $m$ corresponds
to the band mass for GaAs and to the density-dependent
cyclotron mass
\begin{equation}
m = \frac{\hbar k_{\text{F}}(n_0)}{v_{\text{F}}}
\end{equation}
for SLG.
The parameter $C_{\text{q}}$, which comes from the pressure term in the
Navier--Stokes equation, is a capacitance per unit area 
which is given by the following constant value for GaAs
\begin{equation}
C_{\text{q}} = \frac{me^2}{\pi\hbar^2},
\end{equation}
and by
\begin{equation}
 \quad
C_{\text{q}} = \frac{2e^2n_0}{\varepsilon_{\text{F}}(n_0)}
\end{equation}
for SLG.
Here, the density-dependence can be understood as a {\it quantum capacitance} effect~\cite{Yu_2013}, i.e.\ a nonlinear
dependence of the electric potential on the
charge density $-en_0$.

We now eliminate the electric potential from Eq.~\eqref{eq:euler1}
by using the local capacitance approximation
in Eq.~\eqref{eq:localcapac1}, i.e.\ by substituting
\begin{equation}
\varphi_1^{(j)} - \frac{en_1^{(j)}}{C_{\text{q}}}
= \begin{dcases}
- \frac{en_1^{(j)}}{C_j},
& \text{for~odd $j$}, \\
- \frac{en_1^{(j)}}{C_j}
+ \frac{C_{\text{t}}U_{\text{a}}\cos\omega t}{C_{\text{b}}+C_{\text{t}}},
& \text{for~even $j$}, \\
\end{dcases}
\end{equation}
where we have introduced the effective capacitances $C_j$,
\begin{equation}
C_j^{-1} = \begin{cases}
C_1\equiv C_{\text{b}}^{-1} + C_{\text{q}}^{-1},
& \text{odd $j$}, \\
C_2\equiv (C_{\text{b}} + C_{\text{t}})^{-1} + C_{\text{q}}^{-1},
& \text{even $j$}.
\end{cases}
\end{equation}

The linearized hydrodynamic equations describe ``plasma waves''~\cite{Dyakonov_1993, Dyakonov_1995, Dyakonov_1996a,
Dyakonov_1996b, Tomadin_2013, Tomadin_2021},
which, in this formalism and as DS noted, are analogous to surface waves
in shallow water~\cite{Landau:6}.
To find the spatial dispersion of plasma waves we consider the
homogeneous system of equations
\begin{equation}
\begin{pmatrix}
-\I\omega & \I k_x n_0 \\
e^2\I k_x/mC_j &
-\I\omega + \tau^{-1} + \nu k_x^2
\end{pmatrix}
\begin{pmatrix}
n_1^{(j)}(k_x,\omega) \\
v_1^{(j)}(k_x,\omega)
\end{pmatrix}
= 0.
\end{equation}
There are two modes with wave vectors
$\pm K_j(\omega)$, which depend on the region,
with
\begin{equation}
K_j(\omega) =
\sqrt{\frac{\omega^2 + \I\omega/\tau}
{s_j^2 - \I\omega\nu}}.
\end{equation}
Here, $s_j$ is the speed of plasma waves in the $j$-th region,
\begin{equation}
s_j = \sqrt{\frac{eV_{\text{b}}}{m}
\frac{C_{\text{b}}}{C_j}}.
\end{equation}
The first-order solutions are harmonics with the same frequency
$\omega$ of the driving voltage, i.e.
\begin{subequations}
\begin{gather}
n_1^{(j)}(x,t) = n_1^{(j)}(x,\omega) \E^{-\I\omega t} + \text{cc}, \\
v_1^{(j)}(x,t) = v_1^{(j)}(x,\omega) \E^{-\I\omega t} + \text{cc},
\end{gather}
\end{subequations}
where each component is the superposition of two modes:
\begin{subequations}\label{eq:solution1}
\begin{gather}
n_1^{(j)}(x,\omega) = n_0\left(
A_{2j-1}\E^{\I K_j x} + A_{2j}\E^{-\I K_j x}\right), \\
v_1^{(j)}(x,\omega) = \frac{\omega}{K_j}\left(
A_{2j-1}\E^{\I K_j x} - A_{2j}\E^{-\I K_j x}\right).
\end{gather}
\end{subequations}
The $(4N+2)$ coefficients $A_l$ are fixed by the two boundary
conditions in Eq.~\eqref{eq:boundary} and by $4N$ conditions that match the
solutions between any two adjacent regions of the channel.
The two boundary conditions at first order read
\begin{equation}\label{eq:cond1}
\varphi_1^{(1)}(0,\omega) = 0, \quad
v_1^{(2N+1)}(\ell_{2N+1},\omega) = 0.
\end{equation}
A set of $2N$ matching conditions comes from integrating
Eq.~\eqref{eq:cont1} along an infinitesimal interval centered on a
generic point $x$.
It follows that $v_1(x,\omega)$ is continuous in $x$, i.e.
\begin{equation}\label{eq:cond2}
v_1^{(j)}(\ell_j,\omega) = v_1^{(j+1)}(0,\omega).
\end{equation}
The other $2N$ conditions analogously come from Eq.~\eqref{eq:euler1},
which implies the continuity in $x$ of
\begin{equation}
\varphi_1(x,\omega)
- \frac{en_1(x,\omega)}{C_{\text{q}}}
+ \nu \partial_x v_1(x,\omega).
\end{equation}
Rewriting it in terms of $n_1$, the conditions are
\begin{subequations}\label{eq:cond3}
\begin{gather}
\begin{multlined}[b]
\bigg(\frac{C_{\text{b}}}{C_1}
- \frac{\I\omega\nu}{s^2}\bigg)
\frac{n_1^{(j)}(\ell_j,\omega)}{n_0}
= - \frac{C_{\text{t}}}{C_{\text{b}}+C_{\text{t}}}
\frac{U_{\text{a}}}{2V_{\text{b}}} \\
+ \bigg(\frac{C_{\text{b}}}{C_2}
- \frac{\I\omega\nu}{s^2}\bigg)
\frac{n_1^{(j+1)}(0,\omega)}{n_0}, \quad
\text{odd $j$},
\end{multlined} \\
\begin{multlined}[b]
\bigg(\frac{C_{\text{b}}}{C_2}
- \frac{\I\omega\nu}{s^2}\bigg)
\frac{n_1^{(j)}(\ell_j,\omega)}{n_0}
- \frac{C_{\text{t}}}{C_{\text{b}}+C_{\text{t}}}
\frac{U_{\text{a}}}{2V_{\text{b}}} \\
= \bigg(\frac{C_{\text{b}}}{C_1}
- \frac{\I\omega\nu}{s^2}\bigg)
\frac{n_1^{(j+1)}(0,\omega)}{n_0}, \quad
\text{even $j$},
\end{multlined}
\end{gather}
\end{subequations}
where $s^2 = eV_{\text{b}}/m$.
As the frequency $\omega$ is fixed by the antenna,
Eqs.~(\ref{eq:cond1}), (\ref{eq:cond2}), and~(\ref{eq:cond3}) provide a linear system of
($4N+2$) equations for the coefficients $A_l$
appearing in Eq.~\eqref{eq:solution1}.

In the hydrodynamic regime, the electronic mobility has a correction to
the ohmic value $\mu=e\tau/m$, due to the pressure of the electron
fluid.
The mobility relates the electron velocity to the electric field,
$v = \tau\partial_x\varphi$.
Thus, considering the stationary state of Eq.~\eqref{eq:euler1},
the electron mobility assumes different
values between the SGRs and DGRs,
\begin{equation}
\mu_j = \begin{dcases}
\frac{e\tau}{m}
\bigg(1 + \frac{C_{\text{b}}}{C_{\text{q}}}\bigg),
& \text{odd $j$}, \\
\frac{e\tau}{m}
\bigg(1 + \frac{C_{\text{b}}+C_{\text{t}}}{C_{\text{q}}}\bigg),
& \text{even $j$}.
\end{dcases}
\end{equation}

\subsection{Second-order solution: DC photovoltage}
Since we are interested in the calculation of the DC signal
$\Delta U$, we take the time average of the second-order equations.
The local capacitance approximation~\eqref{eq:localcapac} becomes
\begin{equation}
e\delta n^{(j)} = \begin{cases}
-C_{\text{b}}\delta\varphi^{(j)}, & \text{odd $j$}, \\
-(C_{\text{b}}+C_{\text{t}})\delta\varphi^{(j)}, & \text{even $j$}.
\end{cases}
\end{equation}
The time-averaged second-order hydrodynamic equations read
\begin{gather}
\langle \partial_x(n_1v_1) \rangle_t
+ n_0 \partial_x\delta v = 0, \\
\begin{multlined}[b]
\langle v_1\partial_x v_1 \rangle_t
= \frac{e}{m}\partial_x\bigg(
\delta\varphi - \frac{e\delta n}{C_{\text{q}}}\bigg)
- \frac{\delta v}{\tau} + \nu\partial_x^2 \delta v \\
- \eta\bigg\langle
\frac{e}{m}\,\frac{n_1}{n_0}\partial_x \varphi_1
- v_{\text{F}}^2\, \frac{n_1\partial_x n_1}{n_0^2} 
+ \frac{v_1\partial_t n_1}{n_0} \bigg\rangle_{\!t},
\end{multlined}
\end{gather}
where the parameter $\eta$ distinguishes between the two cases
\begin{equation}
\eta = \begin{cases}
0, & \text{GaAs}, \\
\frac12, & \text{SLG}.
\end{cases}
\end{equation}

\begin{figure}
\centering
\includegraphics{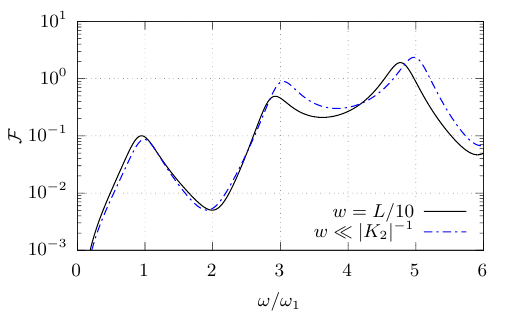}
\caption{%
Dimensionless photovoltage ${\cal F}$, defined in Eq.~(\ref{eq:dlpv}),
for a GaAs FET with a single top gate centered in $x_0=L/2$,
as a function of the frequency $\omega$ of the radiation.
The solid (dash-dotted) line shows the evaluation of Eq.~\eqref{eq:du}
[Eq.~\eqref{eq:fapprox}] for a gate of width $w=L/10$
(vanishing width).
The electron liquid parameters are:
$n_0=\SI{e11}{cm^{-2}}$,
$\tau=\SI{20}{ps}$,
$\nu=0$.
The geometrical parameters are:
$L=\SI{10}{\micro m}$,
$h_{\text{b}}=h_{\text{t}}=\SI{100}{nm}$,
$\varepsilon_{\text{b}}=\varepsilon_{\text{t}}=10$.
With these parameters the frequency $\omega$ shown
on the horizontal axis varies from \num{0}
to $6\omega_1\approx\SI{700}{GHz}$.
}\label{fig:papprox}
\end{figure}

\begin{figure*}
(a) \includegraphics[width=0.45\linewidth]{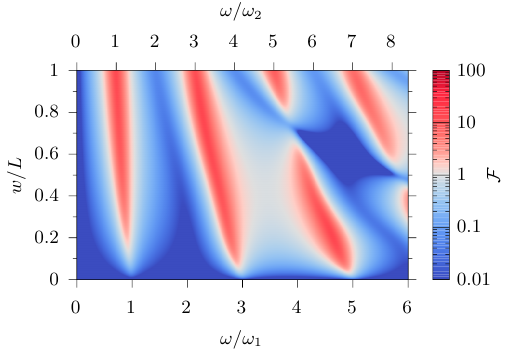}
(b) \includegraphics[width=0.45\linewidth]{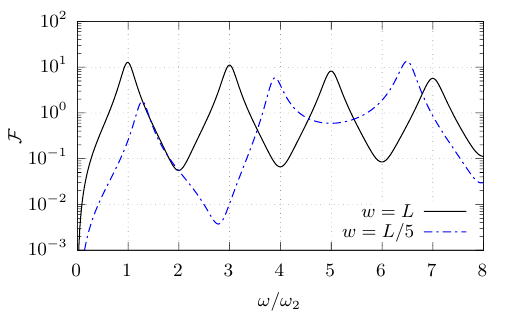}
\caption{%
(a) Dimensionless photovoltage $\mathcal F$ 
for a GaAs FET with a single top gate centered in $x_0 = L/2$,
as a function of the frequency $\omega$
and the gate width $w$.
(b) Horizontal cuts of the profile shown in (a)
for $w=L$ (solid line) and $w=L/5$ (dash-dotted line).
The electron liquid parameters are:
$n_0=\SI{e11}{cm^{-2}}$,
$\tau=\SI{50}{ps}$,
$\nu=\SI{100}{cm^2/s}$.
The geometrical parameters are the same as in Fig.~\ref{fig:papprox}.
These parameters correspond to
the fundamental plasma frequencies
$\omega_1\approx\SI{113}{GHz}$,
$\omega_2\approx\SI{80}{GHz}$.
}\label{fig:pwmap}
\end{figure*}

Taking into account the boundary condition at the drain,
the continuity equation becomes
\begin{equation}
\langle n_1 v_1\rangle_t + n_0 \delta v = 0.
\end{equation}
By using the first-order Eqs.~(\ref{eq:cont1}) and~(\ref{eq:euler1}),
we can rearrange the expression for the DC component as
\begin{multline}
\frac{e}{m}\bigg( \delta\varphi(x)
- \frac{e\delta n(x)}{C_{\text{q}}} \bigg)
= \int_0^x \bigg\langle v_1\partial_x v_1
- (1-\eta)\frac{n_1 v_1}{n_0\tau} + \\
\nu\partial_x^2\frac{n_1v_1}{n_0}
- \eta\frac{v_{\text{F}}^2}{2}\,
\frac{n_1\partial_x n_1}{n_0^2}
- \eta\nu\frac{n_1\partial_x^2 v_1}{n_0}
\bigg\rangle_{\!t} \dif x'.
\end{multline}
The photovoltage $\Delta U$ defined in Eq.~\eqref{eq:dudef}
can be expressed in terms of the first-order
solution in Eq.~\eqref{eq:solution1} as
\begin{multline}\label{eq:du}
\frac{\Delta U}{V_{\text{b}}}
= \sum_j a_j(\omega)\Bigg(
\bigg|\frac{v_1^{(j)}(0,\omega)}{s_1}\bigg|^2
- \bigg|\frac{v_1^{(j)}(\ell_j,\omega)}{s_1}\bigg|^2
\Bigg) \\
+ \sum_j b_j(\omega)\frac{s_j^2}{s_1^2} \Bigg(
\bigg|\frac{n_1^{(j)}(0,\omega)}{n_0}\bigg|^2
- \bigg|\frac{n_1^{(j)}(\ell_j,\omega)}{n_0}\bigg|^2
\Bigg),
\end{multline}
where we have introduced the functions
\begin{gather}
a_j(\omega) = 2 - \eta
- \frac{\nu}{\tau} \bigg(
\frac{1 + \omega^2\tau^2}{s_j^2 + \omega^2\nu\tau}
+ 2 \frac{s_j^2 + \omega^2\nu\tau}{s_j^4 + \omega^2\nu^2}
\bigg), \\
b_j(\omega) = 1
- \eta\bigg(1 + \frac{v_{\text{F}}^2}{2s_j^2}\bigg)
- \omega\nu \frac{\omega\tau - \omega\nu/s_j^2}
{s_j^2 + \omega^2\nu\tau}.
\end{gather}
Finally, since the dependence of the photovoltage on the
amplitude of the incoming radiation is rigorously
second-order because of our perturbative procedure,
it is convenient to introduce a 
dimensionless photovoltage function ${\mathcal F}$
defined by
\begin{equation}\label{eq:dlpv}
\mathcal{F}(\omega) \equiv
\frac{\Delta U(\omega)}{V_{\text{b}}}
\bigg(\frac{U_{\text{a}}}{2V_{\text{b}}}\bigg)^{-2}.
\end{equation}

\subsection{Approximate formula for a single top gate of negligible width}
Let us focus on the case of a single top gate centered in $x_0$,
with negligible width, in order to provide an entirely analytical
result for the photovoltage.
The condition of negligible width translates precisely into
the inequality $|K_{2}(\omega) w|\ll1$.
In order to apply the local capacitance approximation,
we also need $|K_{1,2}(\omega) h_{\text{b,t}}|\ll1$,
and $h_{\text{b,t}}\ll w\ll L$.
For the sake of simplicity we set $\nu=0$.
Calculating the coefficients $A_1,\ldots,A_6$
of the first-order solution in this limit,
we derive the following analytical formula for the photovoltage
\begin{widetext}
\begin{multline}\label{eq:fapprox}
\mathcal F(\omega)
= \bigg[\frac{C_2C_{\text{t}}}{C_{\text{b}}(C_{\text{b}}+C_{\text{t}})}
\frac{\omega w}{s_1}\bigg]^{2}
\frac{1} {\cosh(2\kappa L) + \cos(2kL)}
\bigg\{ (2 - \eta) [\cosh(2\kappa(L-x_0)) + \cos(2k(L-x_0))]
- \bigg[ 1 - \eta\bigg(1 + \frac{v_{\text{F}}^2}{2s_2^2}\bigg) \bigg] \times \\
[\cos(2kx_0)\,\cosh(2\kappa(L-x_0)) +  
\cosh(2\kappa x_0)\,\cos(2k(L-x_0))]
- \bigg[ 1 - \eta\bigg(1 + \frac{v_{\text{F}}^2}{2s_1^2}\bigg) \bigg]
\alpha(\omega)\, [\cosh(2\kappa x_0) - \cos(2kx_0)] \bigg\},
\end{multline}
\end{widetext}
where, for $\nu=0$,
$\alpha(\omega) = \sqrt{1+(\omega\tau)^{-2}}$,
$k(\omega) = \Re K_1(\omega) = \frac{\omega}{s_1}\sqrt{(\alpha(\omega) + 1) / 2}$, and
$\kappa(\omega) = \Im K_1(\omega) = \frac{\omega}{s_1}\sqrt{(\alpha(\omega) - 1) / 2}$.

\begin{figure*}
(a) \includegraphics[width=0.45\linewidth]{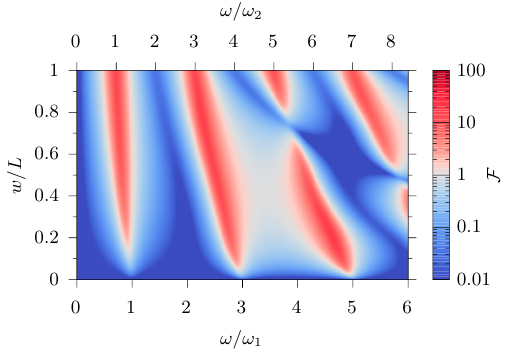}
(b) \includegraphics[width=0.45\linewidth]{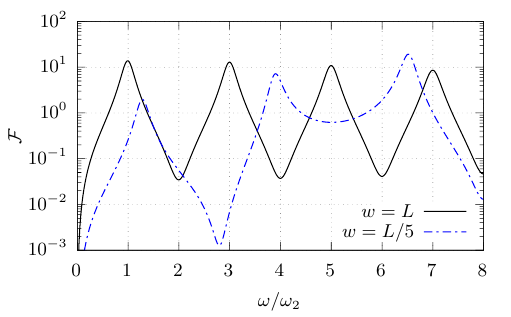}
\caption{%
Same as in Fig.~\ref{fig:pwmap}, but for SLG FET,
with
$n_0=\SI{e12}{cm^{-2}}$,
$\tau=\SI{5}{ps}$,
$\nu=\SI{500}{cm^2/s}$,
$L=\SI{10}{\micro m}$,
$h_{\text{b}}=h_{\text{t}}=\SI{100}{nm}$,
$\varepsilon_{\text{b}}=\varepsilon_{\text{t}}=3$,
corresponding to
$\omega_1\approx\SI{1.16}{THz}$ and
$\omega_2\approx\SI{827}{GHz}$.
}\label{fig:lwmap}
\end{figure*}

\begin{figure*}
(a) \includegraphics[width=0.45\linewidth]{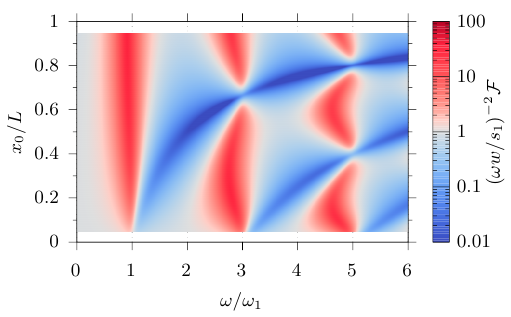}
(b) \includegraphics[width=0.45\linewidth]{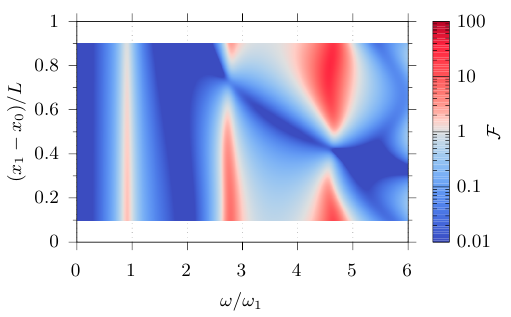}
\caption{%
(a) Dimensionless photovoltage for a GaAs FET
with a single top gate of width $w=L/10$,
as a function of the frequency $\omega$
and the gate position $x_0$.
For the sake of a clearer image
we collect the factor $(\omega w/s_1)^2$.
The parameters are the same of Fig.~\ref{fig:pwmap}.
(b) Dimensionless photovoltage for GaAs,
with two gates of width $w=L/10$,
positioned in $x_0$ and $x_1$ symmetrically
with respect to the center of the channel,
as a function of the frequency $\omega$
and the distance between the two gates.
The other parameters are the same of Fig.~\ref{fig:pwmap}.
}\label{fig:px0map}
\end{figure*}

\begin{figure*}
(a) \includegraphics[width=0.45\linewidth]{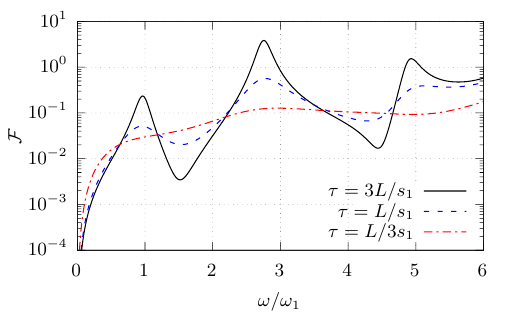}
(b) \includegraphics[width=0.45\linewidth]{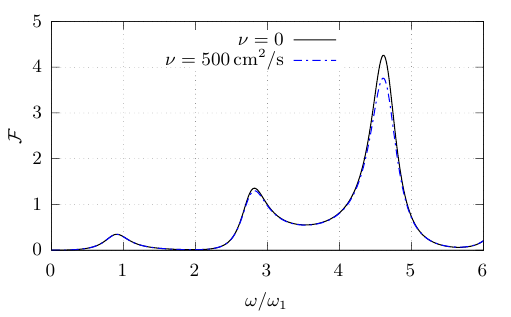}
\caption{%
(a) Dimensionless photovoltage for GaAs,
with a single gate of width $L/10$ centered in $x_0=L/3$,
as a function of the frequency $\omega$,
for several values $\tau=3L/s_1$ (solid line),
$L/s_1$ (dashed line), $L/3s_1$ (dash-dotted line).
The parameters are the same of Fig.~\ref{fig:pwmap},
therefore $L/s_1\approx\SI{14}{ps}$.
(b) Dimensionless photovoltage for SLG,
with a single gate of width $w=L/5$ centered in $x_0=L/2$,
as a function of the frequency $\omega$
for two values of the viscosity $\nu=0$ (solid line)
and $\nu=\SI{500}{cm^2/s}$ (dash-dotted line).
The other parameters are the same of Fig.~\ref{fig:lwmap},
except $\tau=\SI{2}{ps}$.
}\label{fig:lnu}
\end{figure*}

In Fig.~\ref{fig:papprox}, we compare the photovoltage
given by the approximate analytical formula in Eq.~\eqref{eq:fapprox}
to the numerical evaluation of the general formula in Eq.~\eqref{eq:du}.
The approximate formula predicts that the photovoltage has peaks at the
odd harmonics of the fundamental plasma frequency
$\omega_1\equiv\pi s_1/2L$,
as in the original result by DS~\cite{Dyakonov_1996a}.
The numerical results shows that this result is qualitatively
correct, although it also predicts a red-shift of the peak
frequencies on the order of $\sim 10\%$ for the second and third resonances,
and a  sizeable  modulation of the peak amplitude
(please notice the logarithmic scale of the vertical axis).
The analytical formula reported in Eq.~(\ref{eq:fapprox})
can thus be used to obtain a reliable prediction of
the expected photovoltage as a function of the gate position,
if its width is much smaller than the channel.

\section{Geometrical effects on the photovoltage}\label{sec:results}
We now turn to the investigation of the dependence of the photovoltage on
several geometrical parameters of the FET.
Figs.~\ref{fig:pwmap} and~\ref{fig:lwmap} show the profile of the photovoltage
for a device with a single top gate based on GaAs and SLG, respectively.
The color scale represents the amplitude of the dimensionless function ${\cal F}$,
with the radiation frequency $\omega$ and the gate width $w$ (in units of $L$)
on the horizontal and vertical axes, respectively.
The gate width goes from a negligible value (bottom) to the full length
of the channel (top).
Correspondingly, the horizontal axis shows the ratio between the radiation frequency and
the plasma frequency $\omega_{1}\equiv\pi s_1/2L$ of the ungated regions (bottom),
or the plasma frequency $\omega_{2}\equiv\pi s_2/2L$ of the gated region (top).
This double horizontal axis makes it clear that the maxima
of the photovoltage smoothly shift from the odd multiples of $\omega_{1}$
when $w \ll L$ to the odd multiples of $\omega_{2}$ when $w \lesssim L$.
This transition shows that the resonant frequency of the photodetector
crucially depends on the width of the top gate, and that it
is approximately determined by an average of the plasma wave speeds
in the SGRs and DGR.
In the limit in which the top gate occupies the whole length of the
channel, the photovoltage of Eq.~\eqref{eq:du} reproduces
the results of Refs.~\cite{Dyakonov_1996a, Tomadin_2013}.

Fig.~\ref{fig:px0map}(a) shows the photovoltage
for a device based on GaAs with a single top gate of finite width $w=L/10$,
as a function of the frequency $\omega$ on the horizontal axis
and the position $x_0$ of the center of the gate on the vertical axis.
Since the gate has a finite width $w$, the value $x_{0}$ varies between
$w/2$ and $L - w/2$.
The color scale represents the amplitude of the dimensionless function ${\cal F}$
multiplied by a scaling factor to improve visibility of the
profile over the entire parameter range.
This is due to the fact that for a top gate of small width
the photovoltage is proportional to $(\omega w/s_1)^2$,
as we showed in the analytical formula~\eqref{eq:fapprox}.
Familiar maxima of the photovoltage appear at the odd multiples
of the fundamental plasma frequency [cfr.~Fig.~\ref{fig:pwmap}],
but they are intercalated by structures of minimima
which drift in frequency as the position $x_{0}$ of the gate is tuned.
This result shows that a random positioning of the gate
could potentially be very detrimental to the efficiency
of the photodetection device, if it happens to
match one of the minima.

We therefore conclude that a careful calculation of the photovoltage for
given gate position and given width is  necessary prior to fabrication
to optimize the efficiency of the device.
This conclusion, which has been overlooked so far in the previous
theoretical analyses and the experimental fabrication of devices,
is the main result of this Article.
The probability that a top gate is inefficiently positioned is larger
if higher harmonics need to be excited to achieve photodetection,
since it appears that the number of minima in the $n$th harmonic
response profile is $n-1$.
We also notice that tuning the gate position induces a
sensible red-shift of the resonance.
However, since the value of fundamental plasma frequency is
tuned with the {\it bottom} gate by changing
the average electron density, this red-shift can be easily
corrected to achieve frequency-resolved detection.
Fig.~\ref{fig:px0map}(b) shows the dimensionless photovoltage
with $N = 2$ top gates.
Here, for the sake of definiteness,
we choose to consider the simplest extension to the one-gate case,
and we fix the position of the two gates in a symmetric fashion.
We notice that the maxima of the photovoltage still roughly
appear at the odd multiples of the fundamental plasma frequency,
demonstrating the underlying robustness of the wave-mixing
effect at the basis of the DS photodetection scheme.
However, we see substantial modulation of the resonance profile,
both in maximum amplitude and in width.
Moreover, a complex structure of minima intersects the peaks,
once more highlighting the importance of a precise calculation
supporting the device fabrication stage.

Finally, in Fig.~\ref{fig:lnu} we investigate the effects of the
two source of dissipation in the dynamics of the electron liquid,
namely (a) momentum relaxation due to impurity scattering and
(b) electron viscosity.
Fig.~\ref{fig:lnu}(a) shows that the resonance amplitude is
regulated by the dimensionless parameter
$\tau s_1/L$.
Indeed, fixing the device geometry and the electronic density,
the momentum relaxation time $\tau$ determines the resonant response,
since the peaks height grows with the square of the mobility of the sample.
Fig.~\ref{fig:lnu}(b) shows that the peaks at higher harmonics are damped
more strongly due to a finite viscosity.
This is due to the fact that the viscosity term in the Navier--Stokes equation is a second-order
derivative in space, thus proportional to
$\nu K_j^2(\omega)\sim\nu\omega^2/s_j^2$.
In particular, the quality factor is $s_1\tau^*/L$,
where $1/\tau^*=1/\tau+\nu K_1^2$~\cite{Dyakonov_1996a}.
Thus, the quality of the resonance decreses at higher harmonics.

\section{Conclusions}
In this Article, we have presented a generalization of the Dyakonov--Shur
photodetection theory to the case of dual-gated field-effect transistors in which one or
several top gates occupy a limited portion of the channel.
Surprisingly, this configuration, although common in experimental
practice~\cite{Vicarelli_2012, Spirito_2014},
has not been theoretically analyzed so far.
We have provided such analysis and demonstrated that it is much-needed
during the fabrication stage. Indeed, a sub-optimal placement of the
gates can lead to substantially underperforming devices.

We have derived an expression for the photovoltage [Eq.~\eqref{eq:du}]
both for devices where the electron dispersion is parabolic,
as in the case of $n$-doped GaAs quantum wells,
and devices where it is linear, as in the case of single-layer graphene.
Both the width and position of a single top gate affect the way
plasma waves interfere, inducing a modulation of the resonances, as
shown in Figs.~\ref{fig:pwmap} and~\ref{fig:px0map}.
We have also provided a compact analytical formula [Eq.~\eqref{eq:fapprox}]
for the photovoltage, valid in the limit of a negligible single top gate,
which can be used to quickly estimate the optimal gate position.

\begin{acknowledgments}
This work was supported by ``National Centre for HPC, Big Data and Quantum Computing,'' under the National Recovery and Resilience Plan (NRRP), Mission 4 Component 2 Investment 1.4 funded from the European Union -- NextGenerationEU. 
M.P. is supported by the MUR - Italian Ministry of University and Research under the ``Research projects of relevant national interest  - PRIN 2020''  - Project No.~2020JLZ52N (``Light-matter interactions and the collective behavior of quantum 2D materials, q-LIMA'') and by the European Union under grant agreement No. 101131579 - Exqiral and No.~873028 - Hydrotronics. Views and opinions expressed are however those of the author(s) only and do not necessarily reflect those of the European Union or the European Commission. Neither the European Union nor the granting authority can be held responsible for them.
\end{acknowledgments}

\end{document}